\documentclass[journal]{IEEEtran}
\usepackage{amsmath,amsfonts,amsthm,amssymb,mathrsfs}
\usepackage{algorithm}
\usepackage{algpseudocode}
\usepackage{array}
\usepackage{textcomp}
\usepackage{stfloats}
\usepackage{url}
\usepackage{verbatim}
\usepackage{graphicx}
\usepackage{subfigure}
\usepackage{bbding}
\usepackage{multirow}
\usepackage{bm, bbm, xcolor}
\usepackage{refcount}
\usepackage{flafter}
\usepackage{makecell}
\usepackage{diagbox}
\usepackage{booktabs}
\usepackage{graphicx}
\usepackage{cite}
\usepackage{tikz}
\usetikzlibrary{calc}

\theoremstyle{definition}

\newtheorem{remark}{Remark}
\theoremstyle{definition}

\usepackage[T1]{fontenc}
\providecommand{\url}[1]{#1}
\allowdisplaybreaks[4]

\title{Multi-Mode Pinching-Antenna Systems: Mode Selection or Mode Combining?}
\author{
Xiaoxia Xu, %\textit{Member, IEEE}, 
Xidong Mu,  %\textit{Member, IEEE}, 
Yuanwei Liu, \textit{Fellow, IEEE}, 
and Arumugam Nallanathan, \textit{Fellow, IEEE}
\vspace{-0.6cm}
\thanks{X. Xu and A. Nallanathan are with the School of Electronic Engineering and Computer Science, Queen Mary University of
London, London E1 4NS, U.K. (email: \{x.xiaoxia, a.nallanathan\}@qmul.ac.uk).}
\thanks{X. Mu is with the Centre for Wireless Innovation (CWI), Queen's University Belfast, Belfast, BT3 9DT, U.K. (x.mu@qub.ac.uk)}
\thanks{Y. Liu is with the Department of Electrical and Electronic Engineering (EEE), The University of Hong Kong, Hong Kong (e-mail: yuanwei@hku.hk).}
}

\date{\today}

\begin{document}

\maketitle

\begin{abstract}
This letter investigates multi-mode pinching antenna systems (PASS), 
where signals of multiple orthogonal modes can be transmitted within a dielectric waveguide and radiated by pinching antennas (PAs).  
This enables \textit{mode-domain multiplexing} for efficient multi-user communications using a single waveguide.   
In particular, two operating protocols are proposed, namely \textit{mode selection} and \textit{mode combining}. 
Mode selection enforces each PA to predominantly radiate signal power of one single mode, %by matching the propagation constant with this selected mode. 
while mode combining allows each PA to flexibly radiate power of multiple modes. %enables a tunable propagation constant at each PA, thus flexibly radiating power of multiple modes. 
Based on the two protocols, a sum rate maximization problem is formulated for multi-mode PASS-enabled multi-user downlink communications, where the transmit beamforming, PA positions, and PA propagation constants are jointly optimized. 
To address this rapidly oscillating and highly nonconvex problem, a particle swarm optimization (PSO) based 
Karush-Kuhn-Tucker (KKT)-parameterized beamforming (PSO-KPBF) algorithm is proposed. 
KKT-conditioned solutions are exploited to guide the swarm search, thus reducing the search space and achieving fast convergence. 
Numerical results demonstrate that: 
1) Even using a simple uniform mode-combining design, the multi-mode PASS significantly outperform conventional single-mode PASS and hybrid beamforming systems; and  
2) Mode combining achieves high spectral efficiency, while mode selection approximates its performance with a lower hardware complexity. 
Code is released at \url{https://github.com/xiaoxiaxusummer/multi_mode_pinching_antenna}.
\end{abstract}


\section{Introduction}

Pinching-antenna systems (PASS) have recently emerged as a promising wireless technique, where radiating elements, 
termed pinching antennas (PAs), are deployed along dielectric waveguide for near-wired signal transmissions \cite{ref:pass_tutorial_tcom}. 
By leveraging low-loss guided-wave propagation and electromagnetic coupling, PAs can radiate signals from locations near the users, 
thus reducing the large-scale path loss compared to conventional fixed-antenna multiple-input multiple-output (MIMO) systems. 
Recent works have investigated physic modeling and beamforming optimization for PASS \cite{ref:pass_modeling_bf_opt}, 
joint transmit and pinching beamforming \cite{ref:pass_joint_tx_pinching_bf}, 
tri-beamforming design \cite{ref:tribeamforming}, 
adjustable power radiation \cite{ref:power_radiation}, 
line-of-sight (LoS) blockage \cite{los_blockage_PASS}, 
and center-fed PASS designs \cite{ref:cpas_arxiv}. 

Despite these advances, existing PASS with a single mode typically suffer from limited degrees of freedom (DoFs), where each waveguide can only support at most one independent data stream, 
limiting per-waveguide user connections and hardware resource utilization efficiency \cite{ref:pass_survey_arxiv_2026}. 
To overcome this bottleneck, the concept of multi-mode PASS has been proposed \cite{ref:pass_multimode_mu_arxiv}, which utilize multiple guided modes 
to enhance spatial DoF, thus enabling multi-user communications using a single waveguide via mode-domain multiplexing.
However, the efficient exploitation of multi-mode PASS remains largely open. 
In particular, operating protocols to effectively capitalize on the additional DoFs offered by multi-mode PASS are still unclear. 
% The operating protocols to achieve efficient electromagnetic coupling between PAs and multiple modes have not been investigated. 
% Moreover, the tuning of PAs' propagation constants requires digitally programmable structures as well as effective joint design with PAs' locations and transmit beamforming.

In this letter, we propose two fundamental operating protocols for multi-mode PASS, termed \emph{mode selection} and \emph{mode combining}. 
Specifically, mode selection enforces each  PA to select a dedicated guided mode for phase matching, and thus predominantly radiate the signal power of this selected mode. 
In contrast, mode combining customizes the propagation constant for each PA, thereby flexibly radiating signal power of multiple modes. 
For the two proposed protocols, we jointly optimize transmit beamforming, PA positions, and PA propagation constants to maximize the sum rate 
in a multi-mode PASS-enabled multi-user communications.
We develop a particle swarm optimization based \textit{KKT-parameterized beamforming} (PSO-KPBF) algorithm. 
By leveraging the KKT-conditioned beamforming in the swarm search, the proposed PSO-KPBF reduces the search space for fast convergence. 
Simulation results demonstrate that the proposed multi-mode PASS achieves significant gains 
over conventional single-mode PASS and hybrid-beamforming systems for multi-user communications, 
even using a fixed design of uniform mode combining with preconfigured PAs.

\section{System Model and Problem Formulation}
\label{sec:reconfigurable_PA}

\begin{figure}[!t]
    \centering
    \includegraphics[width=0.98\linewidth]{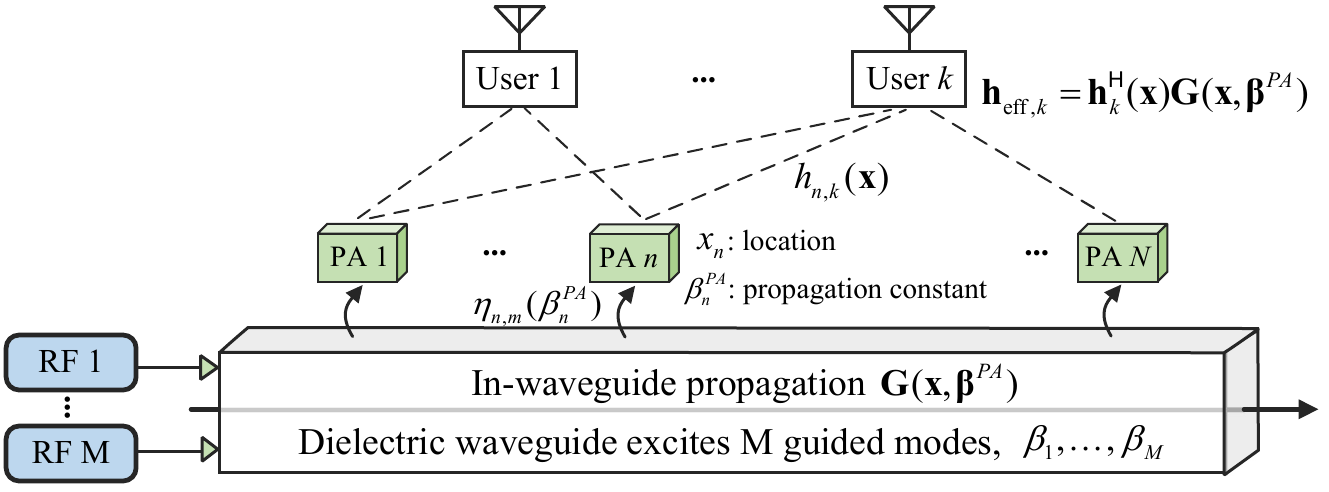}
    \caption{System model of the proposed multi-mode PASS.}
    \label{fig:sys_model}
\end{figure}

% \subsection{System Model}
As shown in Fig. \ref{fig:sys_model}, we consider a downlink PASS that serves $K$ 
single-antenna users by a dielectric waveguide at a fixed height $h_{\mathrm{PA}}$. 
The waveguide excites $M$ orthogonal guided modes, indexed by $\mathcal{M}=\{1,2,\ldots,M\}$, 
obtained from the eigenmodes of Maxwell's equations over the waveguide cross-section. 
These $M$ modes are excited by $M$ radio frequency (RF) chains through $M$ feed points, 
with each RF chain connected to one feed point. 
A set $\mathcal{N}$ of $N$ PAs are placed along the waveguide. 
In a three-dimensional Cartesian coordinate system, the locations of the waveguide entrance and PA $n$ are denoted by $(0,0,h_{\mathrm{PA}})$ and $(x_{n},0,h_{\mathrm{PA}})$. 
Each PA is implemented by an auxiliary short single-mode dielectric waveguide segment, 
whose fundamental mode is characterized by a tunable propagation constant $\beta_n^{\rm PA}$. 
We define $\bm{\beta}^{\mathrm{PA}}\triangleq[\beta_1^{\rm PA},\ldots,\beta_N^{\rm PA}]^{\mathsf T}$ and  $\mathbf{x} \triangleq [x_1,\ldots,x_N]^{\mathsf T}$.

\subsection{Operating Protocols}
\label{subsec:coupling_protocol}
The base station (BS) transmits data symbols $\mathbf s=[s_1,\ldots,s_K]^{\mathsf T}$ with $\mathbb E[|s_k|^2]=1$ 
using a digital precoder $\mathbf W=[\mathbf w_1,\ldots,\mathbf w_K]\in\mathbb C^{M\times K}$. 
Therefore, the signal excited at feed $m$ for mode $m$ is given by $c_{m} = \sum_{k\in\mathcal{K}} w_{m,k} s_k$. 

Let $g_{n,m}$ capture the in-waveguide signal propagation from feed $m$ to PA $n$ carried by mode $m$, 
which is defined as 
\begin{equation}
g_{n,m}(\mathbf x,\bm{\beta}^{\rm PA})
= \eta_{n,m}(\beta_n^{\rm PA}) \, e^{-j\beta_m x_n} \prod_{i=1}^{n-1}\sqrt{1-|\eta_{i,m}(\beta_i^{\rm PA})|^2},
\label{eq:G_cascaded}
\end{equation}
where $\eta_{n,m}(\cdot)$ denotes the electromagnetic (EM) coupling between PA $n$ and mode $m$. 
Hence, $|\eta_{n,m}(\cdot)|^2$ is the power radiation ratio relative to the 
residual power $\prod_{i=1}^{n-1}(1-|\eta_{i,m}(\beta_{i}^{\mathrm{PA}})|^2)$ of mode $m$ before radiating at PA $n$. 
Moreover, $\beta_m=k_0n_{\mathrm{eff},m}$ is the propagation constant of guided mode $m$, 
where $n_{\mathrm{eff},m}$  is the effective refractive index, % obtained by solving the transverse eigenvalue problem subject to the dielectric boundary conditions, 
and $k_0 = \frac{2\pi}{\lambda}$ is the free-space wavenumber with $\lambda$ being the wavelength.

Using coupled-mode theory, $\eta_{n,m}$ can be modelled by \cite{ref:pass_multimode_mu_arxiv}:
\begin{equation}
\eta_{n,m}(\beta_n^{\rm PA})
= \frac{\kappa_{n,m}}{\phi_{n,m}}
\sin(\phi_{n,m}L)
\exp\!\left(-j\frac{L}{2}\Delta\beta_{n,m}\right),
\label{eq:eta_cmt}
\end{equation}
where $\Delta\beta_{n,m} = \beta_n^{PA} - \beta_m$ denotes the phase mismatch between PA $n$ and mode $m$.  
$\phi_{n,m}\triangleq \sqrt{\kappa_{n,m}^2+\Big(\frac{\Delta\beta_{n,m}}{2}\Big)^2}$ denotes the generalized coupling strength, 
$L$ is the length of PA, and  $\kappa_{n,m}$ is the overlap-integral coupling strength.

From \eqref{eq:G_cascaded} and \eqref{eq:eta_cmt}, both the power radiation ratio 
$|\eta_{n,m}(\beta_n^{\rm PA})|^2$ and the resulting phase shift at PA $n$
are determined by the phase mismatch $\Delta\beta_{n,m}$. %, thereby shaping effective channels. 
Motivated by this, we propose two protocols, termed \emph{mode selection} and \emph{mode combining}, 
which configure $\bm{\beta}^{\mathrm{PA}}$  in different manners to alter $\{\Delta \beta_{m,n}\}$ 
and the EM coupling behaviours.    

% To control the sequential power radiation for the signals carried by these $M$ guided modes at each PA, 
% we propose two operating protocols, 
% termed \emph{mode selection} and \emph{mode combining}. 

\subsubsection{Mode Selection}
Each PA acts as a mode-selective coupler, 
which extracts power predominantly from a \emph{single} guided mode for radiation. 
This is achieved by selecting $\beta_n^{\mathrm{PA}}$ from a discrete set:
\begin{equation}
\beta_n^{\mathrm{PA}} \in \mathcal{B}^{\mathrm{MS}} =\{\beta_1,\beta_2,\ldots,\beta_M\},
\label{eq:mode_selection_constraint}
\end{equation}
such that PA $n$ realizes phase matching with one specific mode $m^\star$ and thus $\Delta\beta_{n,m^\star}=0$.
\begin{remark}
The coupling magnitude  $|\eta_{n,m}(\beta_{n}^{\mathrm{PA}})|$ is a monotonically decreasing function of phase mismatch $|\Delta \beta_{n,m}|$. 
Hence, the power radiation ratio $|\eta_{n,m^{*}}(\beta_{n}^{\mathrm{PA}})|^2$ for the selected mode $m^{*}$ 
is maximized at $|\Delta\beta_{n,m^*}|=0$, 
and the maximum value is $|\eta_{n,m^\star}|^2=\big|\sin(|\kappa_{n,m^\star}|L)\big|^2$,
whereas the power radiation ratios for all non-selected modes are suppressed due to non-zero phase mismatch.
\end{remark}

% \textbf{(ii) Mode Combining:}
\subsubsection{Mode Combining}
Unlike mode selection, mode combining enables each PA to tune the propagation constant 
without dedicated phase matching, thus adjusting power radiation ratios of multiple modes. 
Instead of imposing phase matching to any specific $\beta_m$, $\beta_n^{\mathrm{PA}}$ is continuously tunable within a range
\begin{equation}
\beta_n^{\mathrm{PA}} \in \mathcal{B}^{\mathrm{MC}} = [\beta_{\min},\beta_{\max}].
\label{eq:mode_combining_range}
\end{equation}
By appropriately designing phase mismatch $\{\Delta\beta_{n,m}\}$ under specific coupling length $L$, 
\eqref{eq:eta_cmt} can yield multiple large coupling coefficients simultaneously. 
Thus, each PA acts as a programmable multi-mode coupler, 
which can tailor multi-mode power radiation to fully exploit DoFs.  
% for system-level objectives (e.g., sum rate) maximization. 

\begin{remark}
\textbf{(Uniform Mode Combining)} 
A practical and simplified design is the uniform mode combining,  
where all PAs adopt a uniform propagation constant that is fixed at 
\begin{equation}
    \beta_{1}^{\mathrm{PA}} = \beta_{2}^{\mathrm{PA}} = \ldots = \beta_{N}^{\mathrm{PA}} = \frac{\beta_1+\beta_2+\ldots+\beta_M}{2}.
\end{equation}
% which preconfigures coupling strengths for multiple modes.
Our simulation results showcase effectiveness of this design.
\end{remark}

\subsubsection{Hardware Complexity Discussion} 
Mode selection offers discrete control of $\beta_n^{\mathrm{PA}}$ to maximize coupling to a single mode with low hardware complexity,
whereas mode combining provides continuous control of $\beta_n^{\mathrm{PA}}$, 
enabling flexible shaping of multi-mode radiation at the cost of increased hardware complexity. 
Notably, uniform mode combining relies only on preconfigured PAs, 
yielding the simplest implementation but also the least flexibility. 

The physic implementations of mode selection and mode combining are summarized below.
\begin{itemize}
\item \textbf{Mode selection:} 
Mode selection can be realized using corrugated spoof surface plasmon polariton (SSPP) or slow-wave structures, 
whose dispersion (and thus effective propagation constant) is governed by subwavelength corrugation geometry \cite{ref:sspp_pnas}. 
Discrete states of $\beta_n^{\mathrm{PA}}$ can be obtained by preconfigured dispersions. 
PIN diodes embedded in SSPP unit cells enable switching among these states by altering the loading condition. 
Such diode-based reconfigurability has been demonstrated in SSPP-inspired leaky-wave antennas \cite{ref:ssp_lwa_reconfig}.
% and has been demonstrated to be integrated in reconfigurable antennas \cite{ref:reconfig_ant_review}.
\item \textbf{Mode combining:} 
Mode combining relies on tunable materials or reactive loading networks to achieve voltage-controlled modulation of the effective refractive index (or equivalent capacitance/inductance). 
Liquid crystal (LC) technology provides continuous permittivity tuning via voltage-driven molecular reorientation \cite{ref:lc_review_jakoby}. 
When integrated with SSPP structures, strong field confinement enhances dispersion sensitivity to dielectric variation \cite{ref:sspp_pnas}. 
LC-tunable SSPP phase shifters have been experimentally demonstrated \cite{ref:lc_sspp_phase_shifter_ieee}, 
while varactor-loaded spoof-plasmon transmission lines offer an alternative solution \cite{ref:sspp_varactor_admt}.
\end{itemize}

\subsection{Downlink MU-MISO Transmission}
% \subsubsection{Downlink MU-MISO transmission}
Denote the wireless channel from $N$ PAs to user $k$ by $\mathbf h_k(\mathbf x)\in\mathbb C^{N\times 1}$.
Based on spherical-wave assumptions, the line-of-sight (LoS) dominant channel vector is given by
\begin{equation}
    \mathbf{h}_{k}(\mathbf{x})=\left[\frac{\lambda}{4\pi}\frac{e^{-jk_{0}R_{1,k}}}{R_{1,k}},\frac{\lambda}{4\pi}\frac{e^{-jk_{0}R_{2,k}}}{R_{2,k}},\ldots,\frac{\lambda}{4\pi}\frac{e^{-jk_{0}R_{N,k}}}{R_{N,k}}\right]^{\mathsf{T}},
\end{equation}
where $R_{n,k} = \sqrt{x_n^2+h_{\mathrm{PA}}^2}$ denotes the distance between PA $n$ and user $k$.
Denote $\mathbf G(\mathbf x,\bm{\beta}^{\rm PA})\triangleq[g_{n,m}(\mathbf x,\bm{\beta}^{\rm PA})]\in\mathbb C^{N\times M}$. 
The mode-domain effective channel for user $k$ is given by
\begin{equation}
\mathbf{h}_{\mathrm{eff},k}(\mathbf{x},\bm{\beta}^{\rm PA})
\triangleq \mathbf{G}^{\mathsf{H}}(\mathbf x,\bm{\beta}^{\rm PA})\mathbf h_k(\mathbf x)\in\mathbb{C}^{M\times 1},
\label{eq:gk_def}
\end{equation}
and $\mathbf H_{\mathrm{eff}}\triangleq
[\mathbf{h}_{\mathrm{eff},1},\ldots,\mathbf{h}_{\mathrm{eff},K}]\in\mathbb C^{M\times K}$ 
denotes the effective channel matrix. 
The received signal at user $k$ is given by
\begin{equation}
y_k = \mathbf{h}_{\mathrm{eff},k}^{\mathsf{H}}(\mathbf x,\bm{\beta}^{\rm PA})\mathbf w_k s_k
+ \sum_{j\ne k}\mathbf{h}_{\mathrm{eff},k}^{\mathsf{H}}(\mathbf x,\bm{\beta}^{\rm PA})\mathbf w_j s_j
+ n_k,
\label{eq:rx}
\end{equation}
where $n_k\sim\mathcal{CN}(0,\sigma^2)$ is the addictive white Gaussian noise (AWGN). 
The corresponding SINR is
\begin{equation}
\gamma_k(\mathbf W,\mathbf x,\bm{\beta}^{\rm PA})
= \frac{|\mathbf{h}_{\mathrm{eff},k}^{\mathsf{H}}\mathbf w_k|^2}{\sum_{j\ne k}|\mathbf{h}_{\mathrm{eff},k}^{\mathsf{H}}\mathbf w_j|^2+\sigma^2},
\label{eq:sinr}
\end{equation}
and the system sum rate is given by $R(\mathbf W,\mathbf x,\bm{\beta}^{\rm PA})
= \sum_{k=1}^K \log_2\!\big(1+\gamma_k(\mathbf W, \mathbf x,\bm{\beta}^{\rm PA})\big)$.

\subsection{Problem Formulation}
We aim to jointly optimize the beamforming matrix, PA positions, and PA propagation constants:
\begin{subequations}\label{eq:prob_joint}
    \begin{align}
        \max_{\mathbf W,\mathbf x,\bm{\beta}^{\rm PA}}
        \quad & R(\mathbf W,\mathbf x,\bm{\beta}^{\rm PA}) \label{eq:prob_joint_obj}
        \\ \text{s.t.}\quad
        & \|\mathbf W\|_F^2\le P,  \label{eq:con_power}
        \\& x_{n+1}-x_n\ge d_{\min},  ~ \forall n\in\mathcal{N}, \label{eq:x_spacing}
        \\& x_n\in[x_{\min},x_{\max}],  ~ \forall n\in\mathcal{N}, \label{eq:x_range}
        \\& \beta_{n}^{\rm PA}\in\mathcal B^{\mathrm{x}}, ~ \forall n\in\mathcal{N}, ~ \mathrm{x}\in\{\mathrm{MC},\mathrm{MS}\}, \label{eq:prob_joint_cons}
    \end{align}
\end{subequations}
where constraint \eqref{eq:con_power} ensures the maximum transmit power $P_{\max}$. 
\eqref{eq:x_spacing} and \eqref{eq:x_range} enforce the minimum spacing $d_{\min}$ and boundary locations of PAs along the waveguide, respectively. 
Moreover, \eqref{eq:prob_joint_cons} indicates the tunable set of each PA's propagation constant for mode selection or mode combining.  
% with $\mathcal B=\mathcal{B}^{\mathrm{MS}}$ for mode selection  
% and $\mathcal B=\mathcal{B}^{\mathrm{MC}}$ for mode combining. 
% $\mathcal X$ enforces spacing and boundary constraints (e.g., $x_{n+1}-x_n\ge d_{\min}$, $x_n\in[x_{\min},x_{\max}]$)

Problem \eqref{eq:prob_joint} is difficult to tackle due to the highly nonconvex coupled SINR expression in \eqref{eq:sinr} 
and the rapidly oscillatory expression of $\mathbf g_k(\mathbf x,\bm{\beta}^{\rm PA})$ 
in \eqref{eq:G_cascaded}-\eqref{eq:gk_def}. 
To obtain a practically efficient design, we propose a PSO algorithm by integrating the parameterized KKT solutions for beamforming design in the following section.

\section{PSO-KPBF Optimization for Multi-Mode PASS}
\label{sec:sec3_dualbf}

This section develops an efficient joint optimization algorithm. 
We first introduce the \emph{KKT-parameterized beamforming}, which reconstructs  
stationary beamforming solutions from a small set of dual parameters. 
This parameterization substantially reduces the search space and the complexity.
We then propose a model-driven PSO algorithm, namely \emph{PSO-KPBF}, to jointly optimize KPBF parameters, PA locations $\mathbf{x}$, 
and PA propagation constants $\{\beta_n^{\mathrm{PA}}\}$.

%========================================================
\subsection{KKT-Parameterized Beamforming (KPBF)}
\label{subsec:kkt_parbf}

In the proposed multi-mode PASS, the effective downlink channel $\mathbf{H}_{\mathrm{eff}}(\mathbf{x},\bm\beta^{\mathrm{PA}})$ 
depends on both the PA positions $\mathbf{x}$ and the PA propagation constants $\bm\beta^{\mathrm{PA}}$.
Hence, directly optimizing $\mathbf{W}\in\mathbb{C}^{M\times K}$ jointly with $(\mathbf{x},\bm\beta^{\mathrm{PA}})$ leads to 
a strongly coupling nonconvex problem. 
To obtain a tractable design with strong structural prior, we adopt a KKT-inspired parameterization that represents $\mathbf{W}$ 
through a low-dimensional set of nonnegative parameters.
Define a nonnegative vector of dual parameters
$\bm\lambda \triangleq [\lambda_1,\ldots,\lambda_K]^{\mathsf T} \succeq \mathbf{0}$,
and a power allocation vector
$\mathbf{p}_{\mathrm{rel}}\triangleq[p_1,\ldots,p_K]^{\mathsf T}\succeq \mathbf{0}$ satisfying 
$\mathbf{1}^{\mathsf T}\mathbf{p}_{\mathrm{rel}}=1$, 
where $p_k$ specifies the power coefficient allocated to user $k$ to enforce the total power constraint.
Let $\bm\Lambda\triangleq\mathrm{diag}(\bm\lambda)$ and $\mathbf{P}_{\mathrm{rel}}^{1/2}\triangleq\mathrm{diag}(\sqrt{\mathbf{p}_{\mathrm{rel}}})$.
Given $\mathbf{H}_{\mathrm{eff}}\in\mathbb{C}^{K\times M}$ and noise variance $\sigma^2$, we define the unnormalized KPBF precoder as
\begin{equation}
\widetilde{\mathbf{W}}(\bm\lambda,\mathbf{p}_{\mathrm{rel}})
\triangleq
\mathbf{P}_{\mathrm{rel}}^{1/2}\Big(
\mathbf{I}_M
+
\frac{1}{\sigma^2}\mathbf{H}_{\mathrm{eff}}^{\mathsf H}\bm\Lambda \mathbf{H}_{\mathrm{eff}}
\Big)^{-1}
\mathbf{H}_{\mathrm{eff}}^{\mathsf H}.
\label{eq:wtilde_def_twc}
\end{equation}
We then enforce the total transmit power constraint $\mathrm{tr}(\mathbf{W}\mathbf{W}^{\mathsf H})\le P_{\max}$ via a scalar normalization
\begin{equation}
\mathbf{W}(\bm\lambda,\mathbf{p}_{\mathrm{rel}})
\triangleq
\widetilde{\mathbf{W}}(\bm\lambda,\mathbf{p}_{\mathrm{rel}})
\sqrt{\frac{P_{\max}}{\mathrm{tr}\!\left(\widetilde{\mathbf{W}}\widetilde{\mathbf{W}}^{\mathsf H}\right)}},
\label{eq:w_scale_twc}
\end{equation}
which holds with equality whenever $\widetilde{\mathbf{W}}\neq \mathbf{0}$. 
The inverse term in \eqref{eq:wtilde_def_twc} has the form of a weighted regularized Gram matrix. 
Specifically, $\mathbf{H}_{\mathrm{eff}}^{\mathsf H}\bm\Lambda \mathbf{H}_{\mathrm{eff}}$ aggregates user channels with tunable weights $\{\lambda_k\}$, while $\mathbf{P}_{\mathrm{rel}}^{1/2}$ shapes the per-user stream scaling before total-power normalization.
Hence, KPBF searches a structured family of linear precoders with far fewer degrees of freedom than directly optimizing $\mathbf{W}$.

KPBF in \eqref{eq:wtilde_def_twc} can be obtained from KKT conditions of sum-rate maximization 
based on weighted minimum mean square error (WMMSE) reformulation \cite{ref:pass_joint_tx_pinching_bf} by absorbing $\frac{1}{\sigma^2}$ into $\bm{\lambda}$. 
Hence, KPBF can be viewed as a compact, KKT-conditioned parameterization of 
stationary-point beamfoming solutions.

\begin{remark}
Instead of optimizing $\mathbf{W}\in\mathbb{C}^{M\times K}$ directly, 
% KPBF optimizes $\bm\lambda\in\mathbb{R}_+^{K}$ and $\mathbf{p}_{\mathrm{rel}}\in\Delta^{K-1}$, 
% and thus reconstructs $\mathbf{W}$ through \eqref{eq:wtilde_def_twc}-\eqref{eq:w_scale_twc}, 
KPBF reduces to several classical linear precoders as limiting cases.
If $\bm\lambda=\mathbf{0}$ (i.e., $\|\bm\lambda\|\rightarrow 0$),
$\widetilde{\mathbf{W}}=\mathbf{H}_{\mathrm{eff}}^{\mathsf H}\mathbf{P}_{\mathrm{rel}}^{1/2}$,
which corresponds to matched-filter (MRT) precoding with explicit power allocation.
When $\sigma^2\rightarrow 0$ and/or $\bm\lambda$ becomes large so that 
$\frac{1}{\sigma^2}\mathbf{H}_{\mathrm{eff}}^{\mathsf H}\bm\Lambda\mathbf{H}_{\mathrm{eff}}$ dominates, 
the inverse in \eqref{eq:wtilde_def_twc} suppresses multiuser interference, approaching a weighted zero-forcing (ZF) behavior whenever $M\ge K$ and $\mathbf{H}_{\mathrm{eff}}$ has full row rank.
For intermediate $\bm\lambda$ and finite $\sigma^2$, \eqref{eq:wtilde_def_twc} reduces to a weighted MMSE precoder. 
Thus, KPBF searches over a structured WMMSE family through $(\bm\lambda,\mathbf{p}_{\mathrm{rel}})$, 
which can significantly reduce the search space and improve robustness in meta-heuristic search of PSO.
\end{remark}

\subsection{PSO-KPBF Optimization Algorithm}
\label{subsec:pso_kpbf}

The proposed PSO-KPBF algorithm jointly optimizes the PA spatial configuration and the KPBF parameters under a unified swarm-based architecture. 
Specifically, the design variables include: i) the PA locations $\mathbf{x}\!\in\!\mathbb{R}^{N}$ along the waveguide, ii) the PA propagation constants $\bm{\beta}^{\mathrm{PA}}\!\in\!\mathbb{R}^{N}$, and iii) the KPBF parameters $\bm{\lambda}\!\in\!\mathbb{R}_+^{K}$ and $\mathbf{p}_{\mathrm{rel}}\!\in\!\Delta^{K-1}$, where $\bm{\lambda}$ plays the role of nonnegative dual weights and $\mathbf{p}_{\mathrm{rel}}$ specifies the relative power-loading across users. 
Given $(\mathbf{x},\bm{\beta}^{\mathrm{PA}})$, the effective channel $\mathbf{H}_{\mathrm{eff}}(\mathbf{x},\bm{\beta}^{\mathrm{PA}})$ is constructed according to \eqref{eq:gk_def}, and the downlink precoder is reconstructed via \eqref{eq:wtilde_def_twc}-\eqref{eq:w_scale_twc}. 
The swarm then searches the feasible solution space to maximize the downlink sum-rate.

%========================================================
\subsubsection{Discrete and Continuous Optimization of $\bm{\beta}^{\mathrm{PA}}$}
\label{subsubsec:beta_modes}
Depending on the exploited protocols, $\bm{\beta}^{\mathrm{PA}}$ can be optimized in either a discrete-mode form or a continuous form. 
\textbf{i) (Discrete) mode selection.}
For mode selection, $\beta_n^{\mathrm{PA}}\in\{\beta_1,\beta_2\}$ is imposed for all $n\in\mathcal{N}$ in the two-mode case considered in this work. 
We introduce a binary selector vector $\mathbf{b}=[b_1,\ldots,b_N]^{\mathsf T}$ with $b_n\in\{0,1\}$, and define the mapping 
\begin{equation}\label{eq:map_discrete_beta}
    \beta_n^{\mathrm{PA}}=\beta_1(1-b_n)+\beta_2 b_n.
\end{equation} 
The binary variables are updated via binary PSO (BPSO). 
For particle $p$, the binary velocity evolves as
\begin{equation}
v_{b,n}^{(t+1)}
=
\omega v_{b,n}^{(t)}
+
c_1 r_{1,n}^{(t)}
\big(
b_{n,\mathrm{pbest}}^{(t)} - b_n^{(t)}
\big)
+
c_2 r_{2,n}^{(t)}
\big(
b_{n,\mathrm{gbest}}^{(t)} - b_n^{(t)}
\big),
\label{eq:bpso_vel_kpbf}
\end{equation}
followed by stochastic sampling 
\begin{equation}
    b_n^{(t+1)}=\mathbbm{1}\!\left\{u_n^{(t)}<\sigma\!\left(v_{b,n}^{(t+1)}\right)\right\},
\end{equation} 
where $\sigma(x)=1/(1+e^{-x})$ and $u_n^{(t)}\sim\mathcal{U}(0,1)$. 
\textbf{ii) (Continuous) mode combining.}
For mode combining, continuous tuning over $\beta_n^{\mathrm{PA}}\in[\beta_{\min},\beta_{\max}]$ is feasible for each PA $n$. 
The update follows the standard PSO recursion $v_{\beta,n}^{(t+1)}=\omega v_{\beta,n}^{(t)}+c_1 r_{1,n}^{(t)}(\beta_{n,\mathrm{pbest}}^{(t)}-\beta_n^{(t)})+c_2 r_{2,n}^{(t)}(\beta_{n,\mathrm{gbest}}^{(t)}-\beta_n^{(t)})$, and $\beta_n^{(t+1)}=\Pi_{[\beta_{\min},\beta_{\max}]}\!\left(\beta_n^{(t)} + \mathrm{clip}(v_{\beta,n}^{(t+1)})\right)$, where $\mathrm{clip}(\cdot)$ limits the velocity magnitude and $\Pi_{[\beta_{\min},\beta_{\max}]}(\cdot)$ denotes projection onto the feasible interval. 
This provides finer phase control at the expense of a larger search domain.

%========================================================
\subsubsection{Unified PSO-KPBF Architecture}
\label{subsubsec:unified_framework}

The proposed PSO-KPBF embeds the above $\bm{\beta}^{\mathrm{PA}}$ formulations in a common swarm structure. 
Only the update rule of $\bm{\beta}^{\mathrm{PA}}$ differs between the discrete and continuous settings, while all remaining components follow identical PSO dynamics. 
The algorithm employs a swarm of $N_{\mathrm{p}}$ particles to maximize the downlink sum-rate $f(\mathbf{x},\bm{\beta}^{\mathrm{PA}},\bm{\lambda},\mathbf{p}_{\mathrm{rel}})=R\!\left(\mathbf{W}\!\left(\bm{\lambda},\mathbf{p}_{\mathrm{rel}};\mathbf{H}_{\mathrm{eff}}(\mathbf{x},\bm{\beta}^{\mathrm{PA}})\right)\right)$, subject to $x_n \in [x_{\min},x_{\max}]$, $|x_n-x_{n'}|\ge d_{\min}$ for all $n\neq n'$, $\bm{\lambda}\succeq \mathbf{0}$, $\mathbf{p}_{\mathrm{rel}}\succeq \mathbf{0}$, and $\mathbf{1}^{\mathsf T}\mathbf{p}_{\mathrm{rel}}=1$. 

\begin{algorithm}[t]
    \caption{PSO-KPBF for Multi-PA PASS Downlink}
    \label{alg:pso_kpbf}
    \begin{algorithmic}[1]
    \Require Iterations $T$, PSO parameters $N_{\mathrm{p}},\omega,c_1,c_2$.
    \State Initialize particles $\{\bm{\theta}_p^{(0)}\}_{p=1}^{N_{\mathrm{p}}}$ with feasible solutions.
    % \State Evaluate fitness $f_p^{(0)}$ for all particles using \eqref{eq:wtilde_def_twc}-\eqref{eq:w_scale_twc}. 
    \State Set $\bm{\theta}_{p,\mathrm{pbest}}^{(0)}$ and $\bm{\theta}_{\mathrm{gbest}}^{(0)}$.
    \For{$t=0,1,\ldots,T-1$}
        \For{$p=1,\ldots,N_{\mathrm{p}}$}
            \State Update continuous blocks $\{\mathbf{x},\mathbf{z}_\lambda,\mathbf{z}_p\}$ by \eqref{u_update_pso}.
            \State Project $\mathbf{x}\leftarrow \mathcal{P}(\mathbf{x})$, and clamp $(\mathbf{z}_\lambda,\mathbf{z}_p)$.
            % \State Update $\bm{\beta}^{\mathrm{PA}}$:
            \If{$\bm{\beta}^{\mathrm{PA}}$ is discrete}
                \State Update $\mathbf{b}$ by BPSO and map to $\bm{\beta}^{\mathrm{PA}}$ via \eqref{eq:map_discrete_beta}.
            \Else
                \State Update $\bm{\beta}^{\mathrm{PA}}\in[\beta_{\min},\beta_{\max}]$ by standard PSO.
            \EndIf
            \State Reconstruct $\mathbf{W}$ by \eqref{eq:wtilde_def_twc}-\eqref{eq:w_scale_twc} and evaluate $f_p^{(t+1)}$.
            \State Update $\bm{\theta}_{p,\mathrm{pbest}}^{(t+1)}$ and $\bm{\theta}_{\mathrm{gbest}}^{(t+1)}$ if improved.
        \EndFor
    \EndFor
    \State \Return $\bm{\theta}_{\mathrm{gbest}}^{(T)}$ and the corresponding precoder $\mathbf{W}$.
    \end{algorithmic}
\end{algorithm}

% \textbf{Projection/repair operator.} 
% To enforce the PA deployment constraints, we define a projection/repair mapping $\mathcal{P}(\cdot)$ acting on $\mathbf{x}$ as follows. 
% Given a tentative location vector $\tilde{\mathbf{x}}\in\mathbb{R}^{N}$, we first apply element-wise box projection $\tilde{x}_n \leftarrow \Pi_{[x_{\min},x_{\max}]}(\tilde{x}_n)$. 
% We then sort the entries in ascending order (without loss of generality, since PA indexing is arbitrary), yielding $\tilde{\mathbf{x}}_{\mathrm{s}}$. 
% Next, we enforce the minimum-separation constraint by a single forward pass:
% \[
% \tilde{x}_{1}^{\mathrm{s}}\leftarrow \tilde{x}_{1}^{\mathrm{s}},\quad 
% \tilde{x}_{n}^{\mathrm{s}}\leftarrow \max\{\tilde{x}_{n}^{\mathrm{s}},\tilde{x}_{n-1}^{\mathrm{s}}+d_{\min}\},\ \forall n\ge2,
% \]
% followed by a backward pass to ensure the upper bound feasibility:
% \[
% \tilde{x}_{N}^{\mathrm{s}}\leftarrow \min\{\tilde{x}_{N}^{\mathrm{s}},x_{\max}\},\quad 
% \tilde{x}_{n}^{\mathrm{s}}\leftarrow \min\{\tilde{x}_{n}^{\mathrm{s}},\tilde{x}_{n+1}^{\mathrm{s}}-d_{\min}\},\ \forall n\le N-1.
% \]
% Finally, we output $\mathbf{x}=\mathcal{P}(\tilde{\mathbf{x}})$ by reassigning the repaired sorted locations to the particle (the permutation is immaterial). 
% This operator guarantees $\mathbf{x}\in[x_{\min},x_{\max}]^N$ and $|x_n-x_{n'}|\ge d_{\min}$.

\textbf{Projection/repair operator.}
To enforce the PA deployment constraints, we define a projection/repair mapping $\mathcal{P}(\cdot)$ on $\mathbf{x}$.
Given a tentative location vector $\tilde{\mathbf{x}}\in\mathbb{R}^{N}$, we first apply element-wise box projection $\tilde{x}_n \leftarrow \Pi_{[x_{\min},x_{\max}]}(\tilde{x}_n)$.
We then sort the entries in ascending order, yielding $\tilde{\mathbf{x}}_{\mathrm{s}}$.
Next, we enforce the minimum separation via a forward pass $\tilde{x}^{\mathrm{s}}_{n}\leftarrow \max\{\tilde{x}^{\mathrm{s}}_{n},\,\tilde{x}^{\mathrm{s}}_{n-1}+d_{\min}\}$ for $n=2,\ldots,N$, followed by a backward pass $\tilde{x}^{\mathrm{s}}_{n}\leftarrow \min\{\tilde{x}^{\mathrm{s}}_{n},\,\tilde{x}^{\mathrm{s}}_{n+1}-d_{\min}\}$ for $n=N-1,\ldots,1$ with $\tilde{x}^{\mathrm{s}}_{N}\leftarrow \min\{\tilde{x}^{\mathrm{s}}_{N},x_{\max}\}$.
Finally, we set $\mathbf{x}=\mathcal{P}(\tilde{\mathbf{x}})$ by assigning the repaired sorted locations back to the particle (the permutation is immaterial).
By construction, $\mathbf{x}\in[x_{\min},x_{\max}]^{N}$ and $|x_n-x_{n'}|\ge d_{\min}$ for all $n\neq n'$.

\textbf{Particle update.}
To enforce nonnegativity and simplex constraints smoothly, we introduce auxiliary variables $\mathbf{z}_\lambda$ and $\mathbf{z}_p$ 
and set $\bm{\lambda}=\exp(\mathbf{z}_\lambda)$ and $\mathbf{p}_{\mathrm{rel}}=\mathrm{softmax}(\mathbf{z}_p)$, which guarantees $\bm{\lambda}\succeq\mathbf{0}$ and $\mathbf{p}_{\mathrm{rel}}\in\Delta^{K-1}$. 
Accordingly, each particle is represented as $\bm{\theta}=\big(\mathbf{x},\bm{\beta}^{\mathrm{PA}},\mathbf{z}_\lambda,\mathbf{z}_p\big)$, 
where $\bm{\beta}^{\mathrm{PA}}$ is updated by either BPSO or standard PSO as described previously.
At each iteration, the fitness evaluation proceeds as: i) construct $\mathbf{H}_{\mathrm{eff}}(\mathbf{x},\bm{\beta}^{\mathrm{PA}})$; 
ii) reconstruct the KPBF precoder using \eqref{eq:wtilde_def_twc}-\eqref{eq:w_scale_twc}; 
iii) compute the sum-rate objective; and iv) update personal best and global best states. 
For any continuous block $\mathbf{u}\in\{\mathbf{x},\mathbf{z}_\lambda,\mathbf{z}_p\}$, 
the velocity-position update follows 
\begin{equation*}
\mathbf{v}_u^{(t+1)}
=
\omega \mathbf{v}_u^{(t)}
+
c_1 \mathbf{r}_1^{(t)} \odot
(\mathbf{u}_{\mathrm{pbest}}^{(t)}-\mathbf{u}^{(t)})
+
c_2 \mathbf{r}_2^{(t)} \odot
(\mathbf{u}_{\mathrm{gbest}}^{(t)}-\mathbf{u}^{(t)}),
\end{equation*}
\begin{equation}\label{u_update_pso}
\mathbf{u}^{(t+1)}
=
\mathbf{u}^{(t)}
+
\Pi(\mathbf{v}_u^{(t+1)}),
\end{equation}
where $\Pi(\cdot)$ enforces feasibility constraints (using $\mathcal{P}(\cdot)$ for $\mathbf{x}$ and bounded updates for $\mathbf{z}_\lambda,\mathbf{z}_p$).

\textbf{Algorithm \ref{alg:pso_kpbf}} summarizes the entire computation procedure of the PSO-KPBF algorithm. 
% Although global optimality is not guaranteed in general, PSO-KPBF converges to a high-quality solution that are sufficient for stable empirical behavior. 
% Specifically, let $f_{\mathrm{gbest}}^{(t)}$ denote the global-best fitness at iteration $t$. 
% Since the global best is updated only upon improvement, we have
% $f_{\mathrm{gbest}}^{(t+1)} \ge f_{\mathrm{gbest}}^{(t)}$, $\forall t$, 
% and $\{f_{\mathrm{gbest}}^{(t)}\}$ is bounded above for finite $P_{\max}$ and $\sigma^2>0$. 
% Hence, the achieveable sum-rate sequence converges to a certain value. 
The overall computational complexity can be analyzed as follows. 
In each iteration: 
(i) forming $\mathbf{H}_{\mathrm{eff}}$ with complexity $\mathcal{O}(IKM)$; 
(ii) computing $\mathbf{W}$ by \eqref{eq:wtilde_def_twc} requires $\mathcal{O}(KM^2+M^3)$; 
and (iii) evaluating the sum rate scales as $\mathcal{O}(K^2M)$.
Hence, the per-particle per-iteration cost is
$\mathcal{C}_{\mathrm{fit}}
=\mathcal{O}(NKM)
+\mathcal{O}(KM^2+M^3)
+\mathcal{O}(K^2M)$.
With $P$ particles and $T$ iterations, the overall complexity is
$\mathcal{O}\!\left(TP\,\mathcal{C}_{\mathrm{fit}}\right)$.
In the considered multi-mode PASS setting, $M$ is small (e.g., $M=2$), so the dominant runtime typically lies in channel construction 
and $\mathcal{O}(NKM)$ multiplication.

% %==================== Fig. 1: Convergence comparison ====================
% \begin{figure}[!t]
%     \centering
%     \includegraphics[width=0.95\linewidth]{figs/conv_comparison.pdf}
%     \caption{Convergence behavior of different PSO-based designs versus the PSO iteration index.}
%     \label{fig:conv}
% \end{figure}

%==================== Fig. 3: Sum rate versus Pmax ====================
\begin{figure}[!t]
    \centering
    \includegraphics[width=0.95\linewidth]{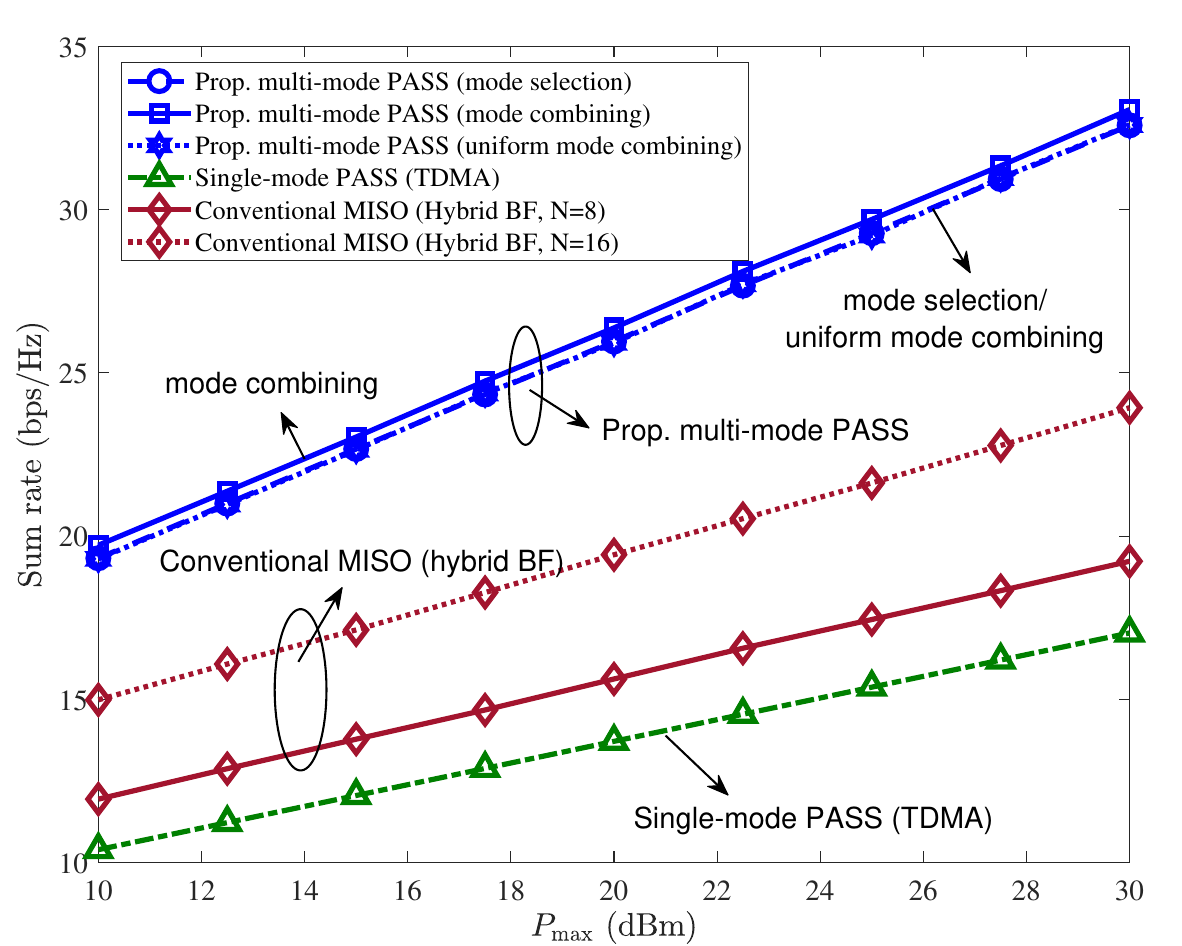}
    \caption{System sum rate versus $P_{\max}$. $N=8$.}
    \label{fig:rate_vs_Pmax}
\end{figure}
%==================== Fig. 2: Sum rate versus N ====================
\begin{figure}[!t]
    \centering
    \includegraphics[width=0.95\linewidth]{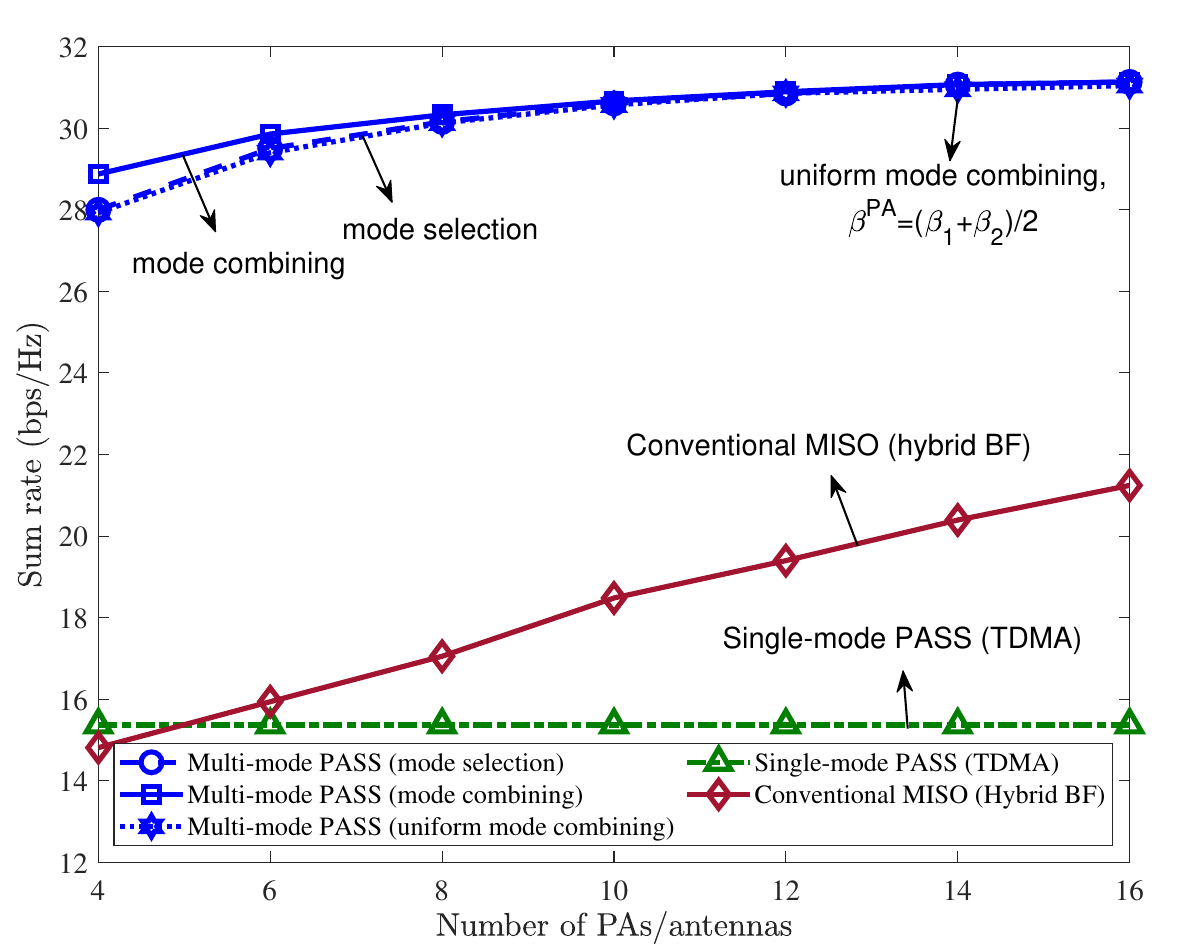}
    \caption{System sum rate versus $N$. $P_{\max}=25$ dBm.}
    \label{fig:rate_vs_N}
\end{figure}

\section{Simulation Results}
\label{sec:simulation}
This section provides numerical results of the proposed multi-mode PASS protocols and the PSO-KPBF algorithm. 
The multi-mode PASS operates at $f=28$ GHz. 
The dielectric waveguide expands $L_{wg}=20$ meters along $x$-axis, which feeds $M=2$ modes through two RF chains to serve $K=2$ users 
for downlink communications. 
The quasi-TE$_0$ and quasi TE$_{1}$ modes are excited, with $\beta_{1}=1009.2378$ rad/m, and $\beta_{2}=645.7996$ rad/m.
The length of each PA is $L=6$ mm\footnote{Detailed settings can be found at \url{https://github.com/xiaoxiaxusummer/multi_mode_pinching_antenna} 
and are omitted here due to space limitations.
}.
We benchmark the multi-mode PASS against both the time-division multiple access (TDMA)-based single-mode PASS and 
conventional hybrid MISO beamforming \cite{ref:hybrid_precoding_jstsp_alkhateeb}.

% Fig.~\ref{fig:conv} illustrates the convergence behavior versus the number of PSO iterations. The proposed PSO-KPBF converges within 
% approximately $50$ iterations and achieves the highest system sum rate. PSO-ZFBF converges to a lower plateau, 
% but outperforms PSO-MRT via interference mitigation. The vanilla PSO without structural beamforming strategy exhibits significantly slower convergence and saturates at a much lower performance level.
% The performance improvement and faster convergence of PSO-KPBF demonstrates its effectiveness in reducing the search space  
% and in guiding the swarm toward high-quality solutions.

Fig.~\ref{fig:rate_vs_Pmax} presents the sum rate versus $P_{\max}$. 
The proposed multi-mode PASS achieve the highest rate, and the performance gap over conventional MISO and single-mode PASS enlarges as transmit power increases.
At low $P_{\max}$, all schemes are primarily noise-limited and exhibit similar scaling behavior. 
As $P_{\max}$ increases, the system becomes interference-limited, and the advantage of structured multi-mode interference management 
becomes more pronounced. The KPBF-based design effectively balances constructive and destructive coupling across modes, 
leading to superior system performance and improved multiplexing efficiency.

Fig.~\ref{fig:rate_vs_N} shows the sum rate versus the number of antennas $N$. 
The sum rates of multi-mode PASS and hybrid beamforming systems increase with $N$ 
due to reduced interference. 
In comparison, single-mode PASS exhibits nearly constant performance due to the limited multiplexing gain. 
Mode combining achieves the highest rate, 
while mode selection and uniform mode combining achieve approximate performance with lower hardwre complexity.
Even the uniform mode combining outperforms hybrid beamforming systems and single-mode PASS, 
demonstrating the efficiency of multi-mode PASS in both path loss reduction and mode-domain multiplexing.

\section{Conclusion}

This letter investigated multi-mode PASS framework for downlink multi-user transmissions.  
Two operating protocols, namely mode selection and mode combining, have been proposed for efficient mode-domain multiplexing. 
Mode selection maximizes the radiated power of a single mode via phase matching, whereas mode combining enables flexible multi-mode power radiation without dedicated phase matching.
% which enables discrete and continuous tuning of PA propagation constants 
% for efficient multi-mode power radiation and mode-domain multiplexing.
The transmit beamforming, PA positions, and PA propagation constants
were jointly optimized to maximize the system sum rate. 
A PSO-KPBF algorithm has been developed to efficiently solve this highly nonconvex problem, which reduced the search space of the particle swarm by parameterized KKT-conditioned beamforming solutions. 
Simulation results demonstrated the performance gains over single-mode PASS and conventional hybrid MISO beamforming, 
which confirmed that the proposed protocols can provide enhanced DoF and mitigate interference for multi-user communications.

%%%%%%%%%%%%%%%%%%%%%%%%%%%%%%%%%%%%%%%%%%%%%%%%%%%%%%%%%%%%%%%%%%%%%%
% % % %%%%%%%%%%%%%%%%%%% - Bibliography Section -  %%%%%%%%%%%%%%%%%%

\bibliographystyle{IEEEtran}

\end{document}